\documentclass[aip,pop,reprint,amsmath,amssymb]{revtex4-1}

\usepackage{makeidx}
\usepackage{graphics}
\usepackage{graphicx,epsfig}
\usepackage{dcolumn}
\usepackage{bm}
\usepackage{rotating}
\usepackage{amsmath}
\usepackage{amssymb}
\usepackage{color}

\begin{document}

\title{A new eddy-viscosity model for large eddy simulation in helical turbulence  }

\author{Changping. Yu$^{1}$
}


\affiliation{$^1$Peking University, Beijing 100871, China\\
            }

\date{\today }
\begin{abstract}
In isotropic helical turbulence, a new single helical model  is
suggested for large eddy simulation. Based on the Kolmogrov's
hypotheses, the helical model is proposed according to the balance
of helicity dissipation and the average of helicity flux across the
inertial range, and the helical model is a kind of eddy viscosity
model. The coefficient of the helical model is constant and can be
also determined dynamically. Numerical simulations of forced and
decaying isotropic helical turbulence demonstrate that the helical
model predicts the energy and helicity evolution well. The
statistical character of the helical model is closer to the DNS
results.In \emph{a priori} test, the energy and helicity
dissipations of the helical model also show a good results. In
general, the helical model has the advantage contrast with the
dynamic Smogorinsky and mixed models.
\end{abstract}

\maketitle

\section{\label{sec1}Introduction}

In turbulence flow, helicity is an important physical quantity. It
is widespread in the motions of the atmosphere, ocean circulation
and other natural phenomena, and also found in leading edge and
trailing vortices shed from wings and slender bodies \cite{a,b,c}.
Helicity can be defined as,
$\emph{h}=\textbf{\emph{u}}\cdot\bm{\omega}$ , where
$\textbf{\emph{u}}$ and $\boldsymbol{\omega}$ are the velocity and
vorticity of the turbulence flow respectively, and \emph{h} is a
pseudoscalar quantity. Similar with the status of energy in the
dynamics of ideal fluids, helicity has the character of inviscid
invariance. This physical property determines that helicity is an
important quantity in turbulence research.

Recently, the researches on helical turbulence have developed
greatly in theoretical research, experiment and numerical
simulation. Based on the helical decomposition of velocity, the
mechanism of existing a joint forward cascade of energy and helicity
has been explained in theory\cite{d}. Cascades existing in helical
turbulence have space scale and time scale\cite{e,f,g,h},and the
researches showed that the existing space scale of helicity cascade
was large than energy cascade. In the inertial range, the joint
cascade of energy and helicity was dominated by the energy cascade
time scale in the low wave number and by the helicity cascade time
scale in high wave number. Using direct numerical simulation of
isotropic turbulence, energy and helicity flux were studied, the
research showed that helicity flux is more intermittent than the
energy flux and the spatial structure were much finer \cite{i}.

Large eddy simulation (LES), as a important method, has been used to
research helical turbulence. Several kinds of SGS models have been
proposed so far, such as eddy-viscosity model, dynamic model, vortex
model, subgrid-scale estimation model, and even the developing
constrained SGS model \cite{j,k,l,m,n,o,p,q,r,s,t,u,v,w}. Although
so many types of SGS model, there are few models on account of the
character of helical turbulence. Y. Li \emph{et al.} proposed a
two-term dynamic mixed models (DSH) to do large eddy simulations of
helical turbulence \cite{x}. Compared the LES results of the DSH
model with other traditional models,the improvement of DSH model was
not so reamarkable.In this paper, we first use the balance of
helicity dissipation and helicity flux across the inertial range to
get the eddy viscosity, then deduce a new helical model (SR). Using
SR model and dynamic SR model (DSR)to do large eddy simulation of
isotropic helical turbulence, we find that SR and DSR models can
predict energy and helicity well. Contrast with dynamic Smogorinsky
model (DSM) and DSH model, the new helical model is a simple model
and its results of LES has get an obvious improvement.

\section{\label{sec2}THEORETICAL ANALYSIS AND SGS MODEL CONSTRUCTION}
In isotropic helical turbulence, the forced helicity control
equation can be get from the  N-S equation, as below
\begin{equation}
\partial_t{h}+\partial_j({u}_j{h})=\partial_j\Omega_j-4\nu{S}_{ij}{R}_{ij}+2{f}_i\omega_i,\label{eq1}
\end{equation}
where ${S}_{ij}=\frac{1}{2}\left(\partial_ju_i+\partial_iu_j\right)$
is the strain rate tensor and
${R}_{ij}=\frac{1}{2}\left(\partial_j\omega_i+\partial_i\omega_j\right)$
is the symmetric vorticity gradient tensor. $\emph{f}_i$ is the
force, and $\Omega_j$ is the flux term, which can be expressed as
\begin{equation}
\Omega_j=-\frac{p}{\rho}+\frac{1}{2}u_iu_i\omega_j+2\nu(
u_iR_{ij}+\omega_iS_{ij})-\varepsilon_{jkl}u_kf_l.\label{eq2}
\end{equation}
For LES, the resolved helicity can be defined as
$h_\Delta=\widetilde{\bm{u}}\cdot\widetilde{\bm{\omega}}$, and we
can get the control function of $h_\Delta$\cite{x} through filtering
the Eq.(1) at scale $\Delta$ as
\begin{equation}
\partial_th_\Delta+\partial_j\left(\widetilde{u}_jh_\Delta\right)=\partial_j\widetilde{\Omega}_j-\Pi_H-4\nu\widetilde{S}_{ij}\widetilde{R}_{ij}+2\widetilde{f}_i\widetilde{\omega}_i,\label{eq2}
\end{equation}
where $\Pi_H=-2\tau_{ij}\widetilde{R}_{ij}$ is the SGS helicity
dissipation rate, and
$\tau_{ij}=\widetilde{u_iu_j}-\widetilde{u}_i\widetilde{u}_j$ is the
SGS stress.

The equation for the resolved helicity in statistically stationary
forced isotropic helical turbulence can be obtained through taking
the ensemble average of Eq.(3) as
\begin{equation}
2\langle\widetilde{f}_i\widetilde{\omega}_i\rangle=-2\langle\tau_{ij}\widetilde{R}_{ij}\rangle+4\nu\langle\widetilde{S}_{ij}\widetilde{R}_{ij}\rangle.\label{eq4}
\end{equation}
where $2\langle\widetilde{f}_i\widetilde{\omega}_i\rangle=\eta$
($\eta$ is the total helicity dissipation)
$-2\langle\tau_{ij}\widetilde{R}_{ij}\rangle$ is the SGS helicity
dissipation, and
$4\nu\langle\widetilde{S}_{ij}\widetilde{R}_{ij}\rangle$ is the
viscous helicity dissipation at scale $\Delta$.

 In the same way, we can also get the equation for the resolved energy in statistically stationary
forced isotropic helical turbulence \cite {y}from the filtered NS
equation as
\begin{equation}
\langle\widetilde{f}_i\widetilde{u}_i\rangle=-\langle\tau_{ij}\widetilde{S}_{ij}\rangle+2\nu\langle\widetilde{S}_{ij}\widetilde{S}_{ij}\rangle.\label{eq5}
\end{equation} where $\langle\widetilde{f}_i\widetilde{u}_i\rangle=\epsilon$
($\epsilon$ is the total helicity dissipation),
$-\langle\tau_{ij}\widetilde{S}_{ij}\rangle$ is the SGS energy
dissipation and
$2\nu\langle\widetilde{S}_{ij}\widetilde{S}_{ij}\rangle$ is the
viscous energy dissipation at scale $\Delta$.

From Eq.(4) and Eq.(5), we can see that the helicity dissipation has
the similar form and composition as the energy dissipation.

In the inertial range of isotropic helical turbulence, the viscosity
of the flow may be ignored, therefor we can get the helicity and
energy relations\cite{z}
\begin{equation}
\eta=-2\langle\tau_{ij}\widetilde{R}_{ij}\rangle,\label{eq6}
\end{equation}
and
\begin{equation}
\epsilon=-\langle\tau_{ij}\widetilde{S}_{ij}\rangle.\label{eq7}
\end{equation}

At the same time, the average of the energy flux across the inertial
range is invariable and equals to the SGS energy dissipation,
\begin{equation}
\langle\Pi_E\rangle=-\langle\tau_{ij}\widetilde{S}_{ij}\rangle,\label{eq8}
\end{equation}
where $\Pi_E$ is the energy flux across the inertial range. Similar
with energy, helicity has the same character in inertial subrange,
\begin{equation}
\langle\Pi_H\rangle=-2\langle\tau_{ij}\widetilde{R}_{ij}\rangle,\label{eq9}
\end{equation}
where $\Pi_H$ is the helicity flux across the inertial range.

In large eddy simulation of isotropic helical turbulence, we take
the eddy-viscosity model as
\begin{equation}
\tau_{ij}-\frac{1}{3}\delta_{ij}\tau_{kk}=-2\nu_T\widetilde{S}_{ij},\label{eq10}
\end{equation}
where $\nu_T$ is the eddy viscosity.

Based on the assumption of Eq.(9) and Eq.(10), we can obtain
\begin{equation}
\langle\Pi_H\rangle=-2\langle\tau_{ij}^{mod}\widetilde{R}_{ij}\rangle=2\cdot\langle2\nu_T\widetilde{S}_{ij}\widetilde{R}_{ij}\rangle,\label{eq11}
\end{equation}
where $\tau_{ij}^{mod}$ is the SGS stress model.

In this deduction, we consider the eddy viscosity $\nu_T$ as a
global averaged value\cite{z1} for the time being, thus
\begin{equation}
\langle\Pi_H\rangle=2\nu_T\langle2\widetilde{S}_{ij}\widetilde{R}_{ij}\rangle.\label{eq12}
\end{equation}

In isotropic helical turbulence, we have
\begin{equation}
\langle2\widetilde{S}_{ij}\widetilde{S}_{ij}\rangle=\int_0^{k_c}2k^2E(k)dk,\label{eq13}
\end{equation}
and
\begin{equation}
2\langle2\widetilde{S}_{ij}\widetilde{R}_{ij}\rangle=\int_0^{k_c}2k^2H(k)dk,\label{eq14}
\end{equation}
where $E(k)$ and $H(k)$ are the energy and helicity spectra
functions, and $k_c$ is the cut off wavenumber.

The energy and helicity spectra $E(k)$ and $H(k)$ have the
inequality\cite{w,x} as
\begin{equation}
|H(k)|\leq2kE(k).\label{eq15}
\end{equation}

In the inertial range, the energy and helicity spectra\cite{y} can
be choosed as
\begin{equation}
E(k)=C_K\epsilon^{2/3}k^{-5/3},\label{eq16}
\end{equation}
and
\begin{equation}
H(k)=C_H\eta\epsilon^{-1/3}k^{-5/3},\label{eq17}
\end{equation}
$C_K$ is Kolmogrov constant and $C_H$ is the coefficient of the
helicity spectrum.

Substituting Eq.(17) into Eq.(14), we can get
\begin{equation}
2\langle2\widetilde{S}_{ij}\widetilde{R}_{ij}\rangle=\frac{3}{2}C_H\eta\epsilon^{-1/3}k_c^{4/3}.\label{eq18}
\end{equation}

From Eq.(15), we can see clearly that
\begin{equation}
\frac{|H(k)|}{E(k)}=\alpha k,\label{eq19}
\end{equation}
where $\alpha$ is a dimensionless parameter, its value changes with
$k$, and $0<\alpha\leq2$.

When the cut wavenumber is in the inertial range, from Eq.(16),
Eq.(17) and Eq.(19), we obtain
\begin{equation}
\epsilon=\frac{C_H}{C_K}\alpha^{-1}k^{-1}|\eta|.\label{eq20}
\end{equation}

Taking the absolute value of both side of Eq.(18), and putting it
into Eq.(20), we can get the expression
\begin{equation}
2|\langle2\widetilde{S}_{ij}\widetilde{R}_{ij}\rangle|=\frac{3}{2}(\alpha
C_KC_H^2)^{1/3}|\eta|^{2/3}k_c^{5/3}.\label{eq21}
\end{equation}
Then, from Eq.(6), (9), (11) and (21), we can obtain the expression
of $\nu_T$,
\begin{equation}
\nu_T=\frac{1}{2}C_r\Delta^{5/2}\widetilde{S}_r,\label{eq22}
\end{equation}
where $\Delta=\pi/k_c$ is the filter scale,
$\widetilde{S}_r=$$|\langle2\widetilde{S}_{ij}\widetilde{R}_{ij}\rangle|^{1/2}$
and the model coefficient $C_r=(\frac{4}{3})^{3/2}\pi^{-5/2}(\alpha
C_KC_H^2)^{-1/2}$. Thus, we obtain a new helical SGS model as
\begin{equation}
\tau_{ij}^{mod}=-C_r\Delta^{5/2}\widetilde{S}_r\widetilde{S}_{ij}.\label{eq23}
\end{equation}

In the LES model, $\alpha$ is set to 2.0 and $C_K=1.6$ and $C_H$ is
confirmed by the DNS data about 1.35.

In this letter, the eddy viscosity $\nu_T$ of the model may be given
a local value,and the expression of $\nu_T$ is
\begin{equation}
\nu_T=\frac{1}{2}C_r\Delta^{5/2}\widetilde{S}_{SR},\label{eq24}
\end{equation}
where
$\widetilde{S}_{SR}=|2\widetilde{S}_{ij}\widetilde{R}_{ij}|^{1/2}$.
Thus we get the new helical model discussed in this letter (SR
model) as
\begin{equation}
\tau_{ij}^{mod}=-C_r\Delta^{5/2}\widetilde{S}_{SR}\widetilde{S}_{ij}.\label{eq25}
\end{equation}
where $C_r\approx0.036$. The coefficient of helical model $C_r$ can
also be decided dynamically, and then we get the DSR model in this
letter.

\section{\label{sec3}THE NUMERICAL RESULTS AND ANALYSIS}
In this part, we will give \emph{a} \emph{priori} and \emph{a
posteriori} test of the LES  model,and do some comparison and
analysis.

In order to validate our model, a DNS of three-dimensional
incompressible homogeneous isotropic turbulence is introduced here.
It solves the forced N-S equations using a pseudo spectral code in a
cubic box with periodic boundary conditions, and the numerical
resolution is $512^{3}$. A Guassian random field is the initial flow
condition, and it has an energy spectrum as
\begin{equation}
E_{0}(k)=A  k^2U_0^2 k_0^{-5}e^{-\frac{2 k^2}{k_0^2}},\label{eq26}
\end{equation}
where $k_{0}=4.5786$  and $U_{0}=0.715$. The whole system is
maintained by a constant energy input rate $\epsilon=0.1$ and
$\eta=0.3$ in the first two wave number shells.

Have constructed the new helical models SR and DSR, we first test
the validity of the models \emph{a priori}. In the inertial range,
we have assumed the invariance of the energy and helicity SGS
dissipations. In such a precondition, the invariance of $f(\delta)$
and $h(\delta)$ must hold,
\begin{equation}
f(\delta)=\langle\delta^{2.5}\widetilde{S}_{SR}\widetilde{S}_{ij}\widetilde{S}_{ij}\rangle,\label{eq27}
\end{equation}
and
\begin{equation}
h(\delta)=\langle\delta^{2.5}\widetilde{S}_{SR}\widetilde{S}_{ij}\widetilde{R}_{ij}\rangle,\label{eq28}
\end{equation}
where $\delta$ is the length scale varying in the inertial range.

\begin{figure}[htbp]
\centering
\epsfig{file=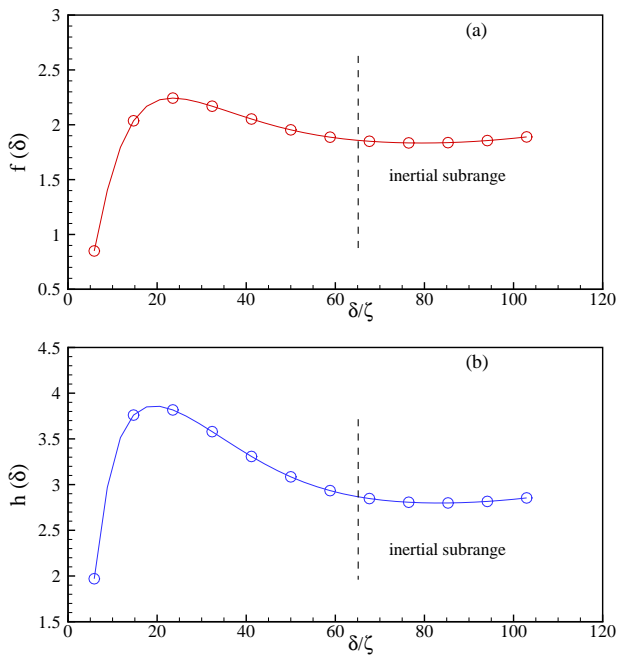,bbllx=85pt,bblly=5pt,bburx=700pt,bbury=600pt,width=0.5\textwidth,clip=}
\caption{(a)$f(\delta)$,(b)$h(\delta)$ distribute with
$\delta/\zeta$ for \emph{a priori}. $\zeta$ is the Kolmogrov length
scale.}\label{fig1}
\end{figure}
In Fig.1, we show $f(\delta)$ and $h(\delta)$ as a function of
$\delta/\zeta$, and $\zeta$ is the Kolmogorov scale.It is easy to
see that the values of the two functions in inertial range are
almost constant, and it demonstrates the scale-invariance and the
reasonableness of the model.

We choose four LES models to do some analysis and comparison here,
the SR model, the DSR model, the dynamic Smogorinsky model (DSM) and
the dynamic mixed helical model (DSH)\cite{u}.

\begin{figure}[htbp]
\centering
\epsfig{file=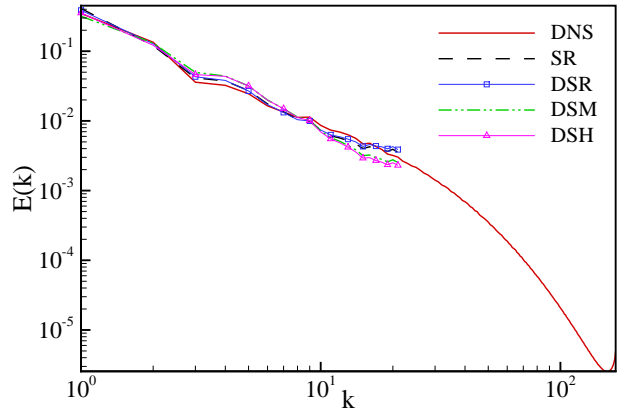,bbllx=85pt,bblly=175pt,bburx=700pt,bbury=550pt,width=0.5\textwidth,clip=}
\caption{Energy spectra for steady isotropic turbulence. Solid line:
DNS; dashed line: SR; line with square: DSR; dash-dot-dotted line:
DSM; line with delta: DSH. }\label{fig2}
\end{figure}
\begin{figure}[htbp]
\centering
\epsfig{file=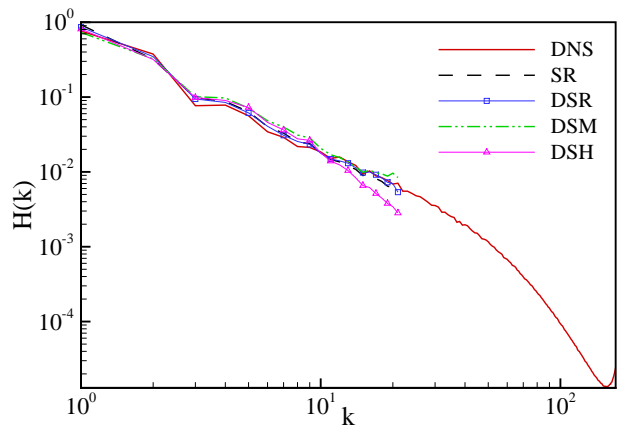,bbllx=85pt,bblly=175pt,bburx=700pt,bbury=558pt,width=0.5\textwidth,clip=}
\caption{Helicity spectra for steady isotropic turbulence. Solid
line: DNS; dashed line: SR; line with square: DSR; dash-dot-dotted
line: DSM; line with delta: DSH.}\label{fig3}
\end{figure}
Fig.2 and Fig.3 display the energy and helicity spectra of DNS and
different models for steady isotopic turbulence respectively. The
bolt solid line is for DNS spectra, and the other lines are for the
four models.It is easy to see that the SR and DSR show similar
trends and predict the energy and helicity spectra better than DSM
and DSH. In Fig.2, DSM and DSH underestimate the energy spectra near
the cutoff wave-number and overestimate at the middle range of the
wave-number.In Fig.3, DSM and DSH overestimate the helicity spectra
at the middle range of the wave-number, and DSH underestimates it
seriously near the cutoff wave-number.

\begin{figure}[htbp]
\centering
\epsfig{file=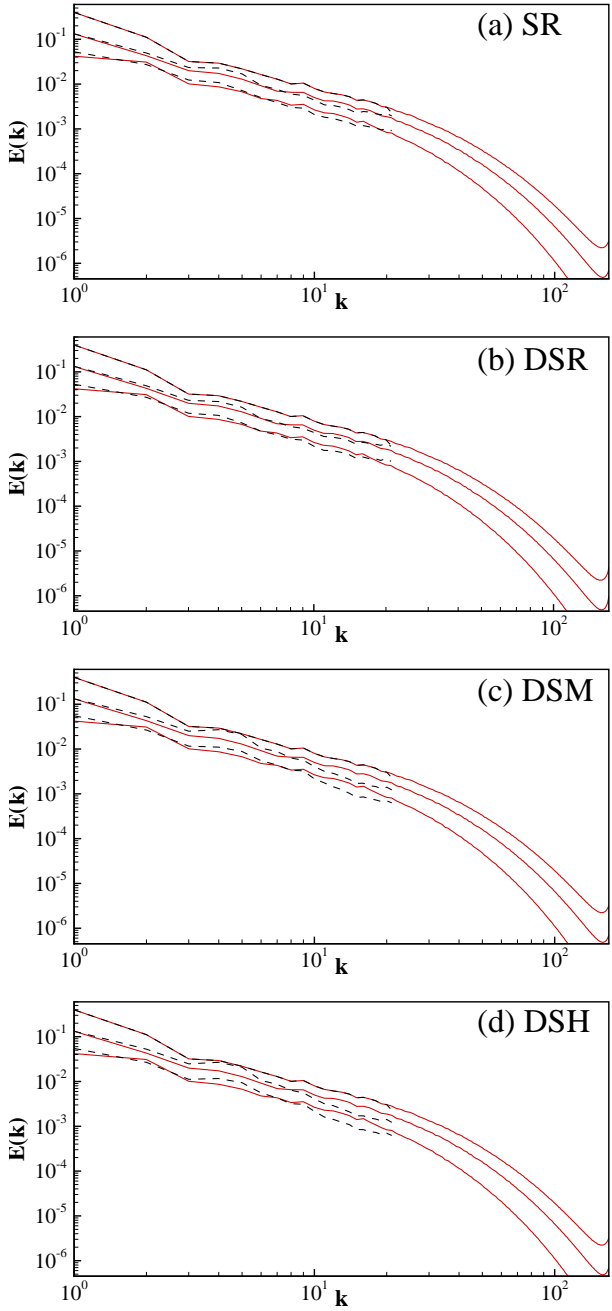,bbllx=85pt,bblly=0pt,bburx=700pt,bbury=605pt,width=1.0\textwidth,clip=}
\caption{Energy spectra for decaying isotropic turbulence ($a
posteriori$), at $t=0, 6\tau_0$, and $12\tau_0$ , where $\tau_0$ is
the initial large eddy turnover time scale. Solid line: DNS; dashed
line: (a) SR, (b) DSR, (c) DSM, (d) DSH .}\label{fig4}
\end{figure}

\begin{figure}[htbp]
\centering
\epsfig{file=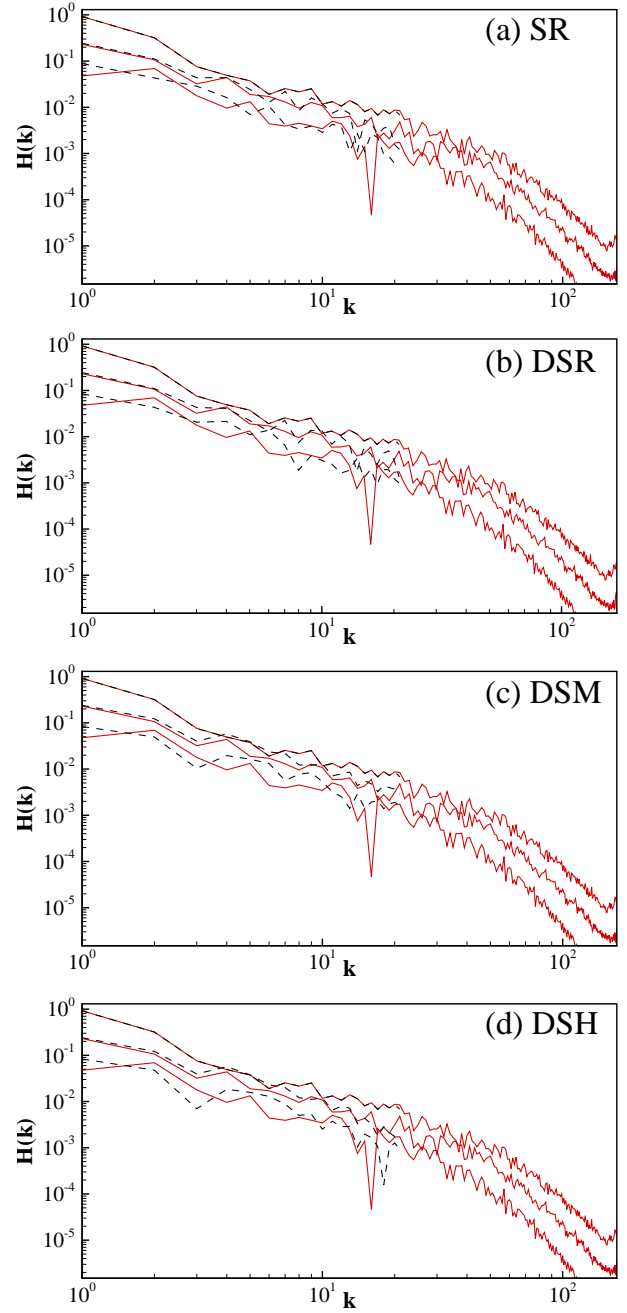,bbllx=85pt,bblly=0pt,bburx=700pt,bbury=605pt,width=1.0\textwidth,clip=}
\caption{Helicity spectra for decaying isotropic turbulence ($a
posteriori$), at $t=0, 6\tau_0$, and $12\tau_0$ , where $\tau_0$ is
the initial large eddy turnover time scale.. Solid line: DNS; dashed
line: (a) SR, (b) DSR, (c) DSM, (d) DSH .}\label{fig5}
\end{figure}
In Fig.4 and Fig.5 we show the time evolution of the DNS and  four
models' energy and helicity spectra for a decay problem starting
from a fully developed statistical steady state respectively, where
$t=0, 6\tau_0$ and $12\tau_0$. In Fig.4, similar with the steady
prbolem, the SR and DSR also show similar trends and predict the
energy spectra very well. DSM and DSH overestimate the energy
spectra at the middle range of the wave-number and underestimate it
near the cutoff wave-number. Fig.5 show us the graphics different
from Fig.3 greatly.The helicity spectra have large fluctuation.
because it is a free decaying course and helicity is a pseudoscalar
quantity , the two factors cause such phenomenon. In Fig.5 we can
still see SR and DSR pridict the helicity spectra a little better
than the other two models.

\begin{figure}[htbp]
\centering
\epsfig{file=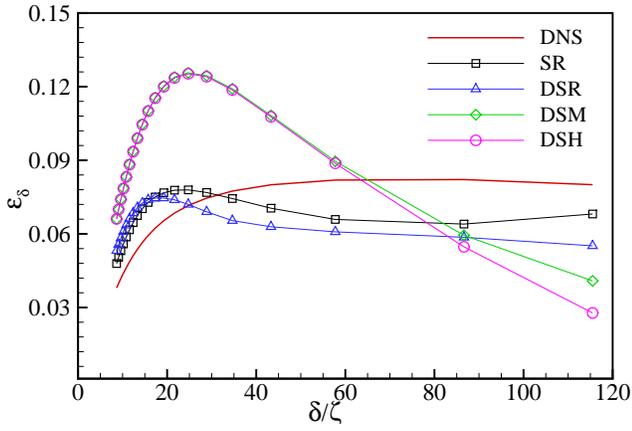,bbllx=85pt,bblly=170pt,bburx=700pt,bbury=555pt,width=0.5\textwidth,clip=}
\caption{The SGS energy dissipation (steady) distribute with
$\delta/\zeta$ for \emph{a priori}. Bold solid line: DNS; line with
Square: SR; line with delta: DSR; line with diamond: DSM; line with
circle: DSH. }\label{fig6}
\end{figure}

\begin{figure}[htbp]
\centering
\epsfig{file=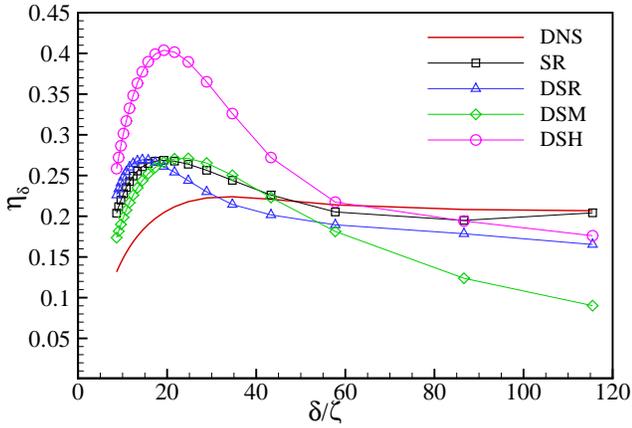,bbllx=85pt,bblly=170pt,bburx=700pt,bbury=555pt,width=0.5\textwidth,clip=}
\caption{The SGS helicity dissipation (steady) distribute with
$\delta/\zeta$ for \emph{a priori}. Bold solid line: DNS; line with
Square: SR; line with delta: DSR; line with diamond: DSM; line with
circle: DSH.}\label{fig7}
\end{figure}
Fig.6 and Fig.7 show us the SGS energy and helicity dissipations of
DNS and the four models distribute with $\delta/\zeta$ for $a$
$priori$ respectively, and they reflect case of the full developed
steady turbulence. It is obvious from in the inertial range that the
SGS energy and helicity dissipations from SR are closer to those of
the DNS than other models, and specially have the similar change
trends with DNS.

\begin{figure}[htbp]
\centering
\epsfig{file=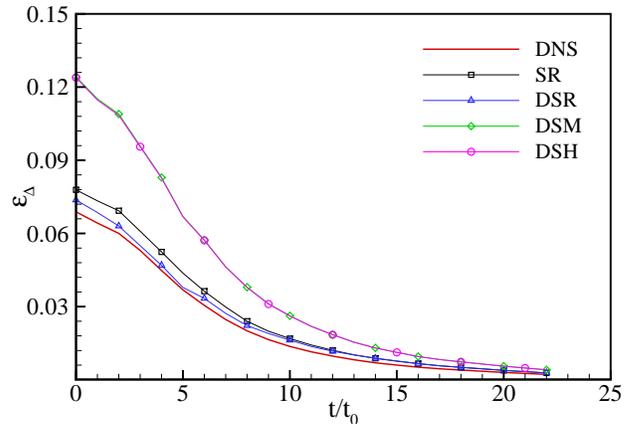,bbllx=85pt,bblly=170pt,bburx=700pt,bbury=555pt,width=0.5\textwidth,clip=}
\caption{The decay of SGS energy dissipation with $t/\tau_0$
($\tau_0$ is the initial large eddy turnover time scale) from a
fully developed steady state for $\emph{a priori}$ at the filter
scale $\Delta$ . Bold solid line: DNS; line with Square: SR; line
with delta: DSR; line with diamond: DSM; line with circle:
DSH.}\label{fig8}
\end{figure}

\begin{figure}[htbp]
\centering
\epsfig{file=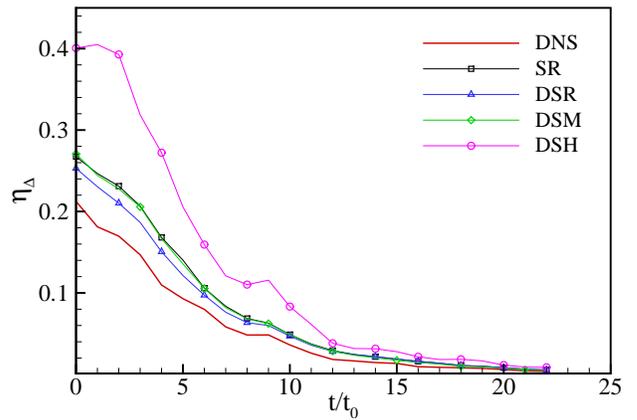,bbllx=85pt,bblly=170pt,bburx=700pt,bbury=555pt,width=0.5\textwidth,clip=}
\caption{The decay of SGS helicity dissipation with $t/\tau_0$ from
a fully developed steady state for $\emph{a priori}$ at the filter
scale $\Delta$ . Bold solid line: DNS; line with Square: SR; line
with delta: DSR; line with diamond: DSM; line with circle:
DSH.}\label{fig9}
\end{figure}
In Fig.8 and Fig.9, we show the decay of SGS energy and helicity
dissipation with $t/\tau_0$($\tau_0$ is the initial large eddy
turnover time scale) from a fully developed steady state for
$\emph{a priori}$ at the filter scale $\Delta$ respectively. It is
clear in Fig.8 that the SGS energy dissipations from SR and DSR are
closer to the DNS result than DSM and DSH, and also have similar
change trend. And also we can see in Fig.9 that the SGS helicity
dissipations from SR, DSR and DSM are closer to the DNS result than
DSH.
\begin{figure}[htbp]
\centering
\epsfig{file=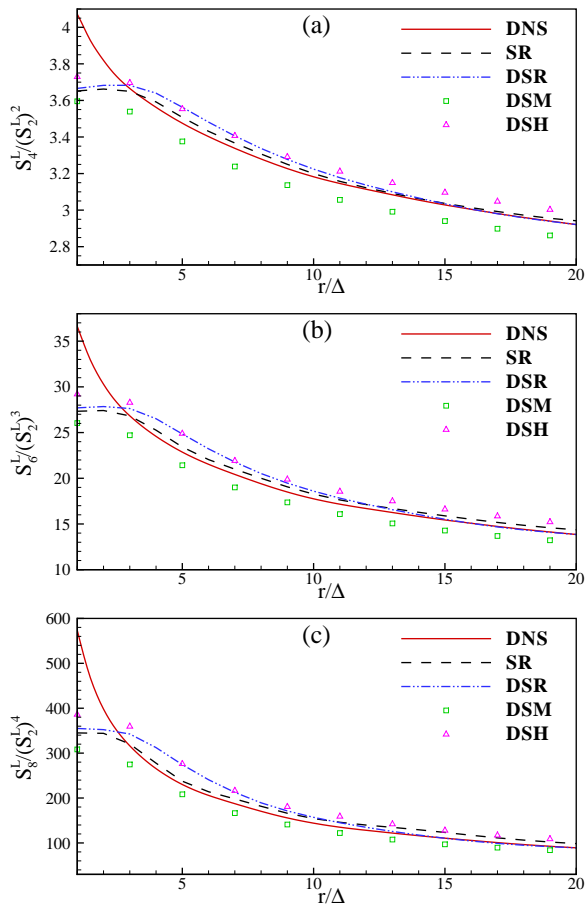,bbllx=85pt,bblly=10pt,bburx=700pt,bbury=600pt,width=0.7\textwidth,clip=}
\caption{The distribution of high-order moments of longitudinal
velocity increment with the separation distance $r$, where $\Delta$
is the filter scale. (a) $S_4^L$, (b) $S_6^L$, (c) $S_8^L$
}\label{fig10}
\end{figure}

In Fig.10,we display the distribution of high-order velocity
increment with $\Delta$, where $r$ is the separation distance and
$\Delta$ is the filter scale. Fig.10 (a), (b) and (c) denote the
fourth-order, sixth-order and eighth-order structure functions of
the longitudinal velocity increment of DNS and the four models
distribute with separation distance. We can from Fig.10 that the
result from SR is closest to the DNS result very well in the range
of $r/\Delta\geq3$, and the behavior of DSR is a little better than
DSH and DSM. While in the range of $r/\Delta<3$, anyone of the
models have still not given a rather good result.

\section{\label{sec4}CONCLUSIONS}
In this letter, we construct a new helical SGS stress model for
large eddy simulation of the isotropic helical turbulence. Different
other eddy viscosity model, we conduct the helical model based on
the balance of helicity dissipation and the average of helicity flux
across the inertial range in helical turbulence. Then we have tested
the scale-invariance of the model for $a$ $priori$ in inertial
range, and make sure the validity of the helical model. We have
given a constant coefficient helical model and a dynamic helical
model here.

Have tested from $a$ $priori$ and $a$ $postoriori$, the helical
model is confirmed to predict energy and helicity spectra
accurately, and also gives rather good simulating results in energy
and helicity dissipation, $et$ $al$.Besides the advantages above
all, the helical model is a single model and can improve the
computing efficiency greatly. As a SGS model on account of helical
turbulence, it can also apply to the rotational turbulence.

\section{\label{sec5}acknowledgement}
This work was supported by the National Natural Science Foundation
of China (Grant Nos. !!!!!!!!!).


\begin{thebibliography}{aipsamp}
\bibitem{a}D. K. Lilly,  \emph{The development and maintence of rotation inconvective storms}
In \emph{Intense atmospheric vortices: proceedings of the joint
symposium (IUTAM/IUGG) held at Reading}, edited by L. Bengtsson and
J. Lighthill (Springer-Verlag, Beilin, 1982),  pp. 149-60.




\bibitem{b}A. Tsinober and E. Levich,  ``On the helical nature of three
dimensional coherent structures in turbulent flows," Phys. Rev.
Lett. \textbf{99}A, 321 (1983).
\bibitem{c} H. K. Moffatt and A. Tsinober, ``Helicity in laminar and turbulent flow," Annu. Rev. Fluid Mech. \textbf{24}, 281
(1992).
\bibitem{d}Q. Chen, S. Chen, and G. L. Eyink, ``The joint cascade of
energy and helicity in three- dimensional turbulence," Phys. Fluids
\textbf{15}, 361 (2003).
\bibitem{e}P. D. Diltevsen and P. Giuliani, ``Cascades in helical
turbulence," Phys. Rev. E \textbf{63}, 036304 (2001).
\bibitem{f}P. D. Ditlevsen and P. Giuliani, ``Dissipation
in helical turbulence," Phys. Fluids \textbf{13}, 3508 (2001).
\bibitem{g}R. H. Kraichnan, ``Inertial-range transfer in two-and three-dimensional turbulence," J. Fluid Mech. \textbf{47},
525 (1971).
\bibitem{h}S. Kurien, M. A. Taylor, and T. Matsumoto, ``Cascade time scales for energy and helicity in
homogeneous isotropic turbulence," Phys. Rev. E \textbf{69}, 066313
(2004).
\bibitem{i} Q. N. Chen, S. Y. Chen, G. L.
Eyink, and D. D. Holm, ``Intermittency in the joint cascade of
energy and helicity," Phys. Rev. Lett. \textbf{90}, 214503 (2003).
\bibitem{j} J. W. Deardoff,
" The use of subgrid transport equation in a three dimensional model
of atmospheric turbulence," ASME. \textbf{26}(6), 669 (1973).
\bibitem{k}S. B. Pope, \emph{Turbulence flow} (Cambridge University Press, Cambridge, 2000)
\bibitem{l}J. Smagorinsky, ``General circulation experiments with primitive equation," Monthly
Weather Review, \textbf{91}, 99 (1963).
\bibitem{m} P. Moin and J. Kim, ``Numerical investigation of turbulent channel flow,"
J. Fluid Mech. \textbf{118}, 341 (1982).
\bibitem{n}M. Germano, U. Piomelli, P. Moin, and W. Cabot, ``A dynamic
subgridscale eddy viscosity model," Phys. Fluids A \textbf{3}, 1760
(1991).
\bibitem{o} D. K. Lilly, ``A proposed modification of the Germano subgrid-scale
closure  method," Phys. Fluids A, \textbf{4}, 3 (1992).
\bibitem{p}A. Misra and D. I. Pullin, ``A vortex-based subgrid
stress model for large eddy simulation," Phys. Fluids \textbf{9},
2443 (1997).
\bibitem{q}J. A. Domaradzki and E. M. Saiki, ``A subgrid-scale model based on
the estimation of unresolved scales of turbulence," \textbf{9}, 2148
(1997).
\bibitem{r}J. A. Langford and R. D. Moser, ``Optimal LES
formulations for isotropic turbulence," J. Fluid Mech. \textbf{398},
321 (1999).
\bibitem{s}Y. Mornishi and O. V. Vasilyev, ``A recommended
modification to the dynamic two-parameter mixed subgrid scale model
for large eddy simulation of wall bounded turbulent flow," Phys.
Fluids \textbf{13}, 3400 (2001).
\bibitem{t}C. Meneveau, ``Statistics of turbulence
subgrid-scale stresses; Necessary conditions and experimental
tests," Phys. Fluids \textbf{6}, 815 (1994).
\bibitem{u}S. Ghosal, T. S. Lund, P. Moin,
and K. Akselvoll, ``A dynamic localization model for large-eddy
simulation of turbulent flows," J. Fluid Mech. \textbf{286}, 229
(1995).
\bibitem{v}Y. Li and C. Meneveau,
``Analysis of mean momentum flux in subgrid models of turbulence,"
Phys. Fluids \textbf{6}, 3483 (2004).
\bibitem{w}Y. Shi, Z. Xiao, and S. Chen,``Costrained subgrid-scale stress model for large sddy simulation,"
Phys. Fluid \textbf{20}, 011701 (2008).
\bibitem{x}Y. Li, C. Meneveau, S. Chen, and G. L. Eyink,
``Subgrid-scale modeling of helicity and energy dissipation in
helical turbulence," Phys. Rev. E \textbf{74}, 026 (2006).
\bibitem{y}C. Meneveaua and T. S. Lund, ``The dynamic Smagorinsky model and scale-dependent coefficients
in the viscous range of turbulence," Phys. Fluid \textbf{9}, 3932
(1997).
\bibitem{z}D. k. Lilly, " The representation of small-scale
turbulence in numerical simulation experiments," \emph{Proceedings
of the IBM Scientific Computing Symposium on Environmental
Sciences}, 1967, p. 195.
\bibitem{z1}S. Cerutti and C. Meneveau, ``Intermittency and relative scaling of subgrid-scale energy dissipation in isotropic turbulence,"
Phys. Fluid \textbf{10}, 928 (1998).
\bibitem{z2}A. Brissaud, U. Frisch, J. Leorat, M. Lesieur, and A. Mazure, "
helicity cascades in fully developed isotropic turbulence," Phys.
Fluids \textbf{16}, 1366 (1973)
\bibitem{z3}V. Borue and S. A. Orszag, ``spectra in helical
three-dimensional homogeneous isotropic turbulence," Phys. Rev. E
\textbf{55}, 7005 (1997).
\bibitem{z4}Y. Zang, R. L.
Street, and J. R. Koseff, ``A dynamic mixed subgrid-scale model and
its application to turbulent recirculating flows," Phys. Fluids A
\textbf{5}, 3186 (1993).
\bibitem{z5}S. W. Liu, C.
Meneveau, and J. Katz, ``On properties of similarity subgrid-scale
models as deduced from measurements in a turbulent jet," J. Fluid
Mech. \textbf{275}, 83 (1994).
\bibitem{z6}C. Meneveau and J. Katz, ``Scale-Invariance and turbulence models for large-
eddy simulation,"  Annu. Rev. Fluid Mech. \textbf{32}, 1 (2000).
\bibitem{z7}H. Lu, C. J. Rutland, and L. M. Smith, ``A priori tests of one-equation LES
modeling of rotating turbulence," J. Turbulence. \textbf{8}, 37
(2007).




\bibitem{z10}Q. Chen and G. L. Eyink, ``The joint cascade of energy and helicity in three-dimensional turbulence," Phys.
Fluid \textbf{15}, 361 (2003).
\bibitem{z11}C. Meneveau and J. Katz, ``Scale-invariance and turbulence models for large-eddy simulation,"
Annu. Rev. Fluid Mech. \textbf{32}, 1 (2000).





\end{thebibliography}
\end{document}